\documentstyle[12pt]{article}
\title{
Computer Simulation of the Tevatron Crystal Extraction Experiment}
\author{Valery Biryukov\thanks{E-mail:  biryukov@mx.ihep.su  }
\\ \vspace{5mm} IHEP Protvino, 142284 Moscow Region, Russia 
\\ \vspace{10mm} Published: {\it in *Dallas 1995, PAC, vol. 3* pp.1958-1960}}
\date{May 1-5, 1995}
\begin{document}
\maketitle

\abstract{
The Fermilab crystal-extraction experiment E853 at Tevatron
was simulated by Monte Carlo code CATCH [1]
tested earlier in the CERN-SPS experiment [2--4].
Predictions for the extraction efficiency, angular scans
and extracted beam profiles are presented.
Several ideas are proposed and tested by the simulation,
how to get in E853
the key information of the extraction experiment:
the "septum width" of a crystal and dependence of extraction
efficiency on it,  the impact parameters of protons at crystal,
and the contribution of the first and multi passes to the extraction.
The ways to optimize E853 are analyzed.
}

\section{Introduction}

The crystal-extraction experiments
at CERN SPS [2--4] and Fermilab Tevatron [5]
have in view possible application of channeling
for proton extraction from a multi-TeV machine [6].
The technique employs a bent crystal placed in the beam halo,
which traps and bends the particles parallel to the crystallographic
plane within Lindhard angle $\theta_c$.
The halo particles
hit a crystal very close to its edge,
with impact parameter $b$ in the range $\sim$\AA{} to $\sim$$\mu$m.
This calls for a good perfection of the crystal edge.
Alternatively,
one should investigate how crystal extracts particles
in the multipass mode, which involves
several scatterings in the crystal of the circulating particles.

As the extraction includes many passes,
there is no easy way to extrapolate the results
with energy.
This makes the detailed comparison of the measurements
with computer simulation essential.
Such an analysis made [7]
for the CERN-SPS experiment
has shown good agreement of the theory with measurements [2,3].
The major result of [7] was a prediction of the edge
imperfection of the crystals used at SPS.
The new SPS experiment, employing a crystal with
an amorphous edge-layer to testify this idea,
has proved much the same efficiency indeed [4].
Another prediction, made for the "U-shaped" crystal
-- much the same efficiency but narrow (70 $\mu$rad {\em fwhm})
angular scan[7] --
has also been confirmed [3].
With the simulation code [1] tested
at SPS, here we model the extraction of 900-GeV protons from
Tevatron, matching E853 [5,8].

The real crystal has an irregularity of the surface,
which defines a range of inefficient $b$ at the edge
("septum width" $t$).
The following information is essential for understanding
the crystal extraction process:

(a) efficiency $F$, and contributions to it from the first/multi passes;
(b) distribution over $b$ at the crystal;
(c) septum width $t$;
(d) dependence $F(t)$.

We propose the ways to get this information in E853.

\section{Qualitative discussion}

The essential feature of E853
is the fact that
the crystal atomic planes are  perpendicular
to the crystal face in touch with the beam.
In E853
one should align a crystal in 2 planes:
the channeling plane (vertical, $y$)
with the accuracy of $\theta_c$,
and the horizontal plane ($x$) to keep the crystal face
parallel to the incident protons (Fig. 1).
At first glance, the need to tune two angles
is an inconvenience.
Here we show that this extra degree of freedom is an excellent
possibility to study  extraction
in many details!
\begin{figure}[bth]
\begin{center}
\setlength{\unitlength}{0.55mm}\thicklines
\begin{picture}(70,82)(-10,6)

\put(0,95){\line(1,0){50}}
\put(50,95){\line(0,-1){20}}
\put(0,95){\line(0,-1){20}}
\put(-7,100){\vector(1,0){50}}
\put(-7,98){\vector(1,0){50}}
\put(-7,93){\vector(1,0){15}}
\put(-7,85){\vector(1,0){15}}
\multiput(0,90)(5,0){10}{\line(1,0){3}}
\multiput(2,95)(5,0){10}{\line(1,-1){4}}
\put(54,92.5){\vector(0,1){2.5}}
\put(54,92.5){\vector(0,-1){2.5}}
\put(56,92){t}

\put(-7,64){\vector(1,0){50}}
\put(-7,62){\vector(1,0){50}}
\put(-7,56){\vector(1,0){15}}
\put(-7,49){\vector(1,0){15}}
\put(0,52){\line(6,1){48}}
\put(0,52){\line(1,-6){3}}
\put(48,60){\line(1,-6){3}}
\multiput(0,52)(4.8,0){10}{\line(1,0){3}}
\put(54,56){\vector(0,1){4}}
\put(54,56){\vector(0,-1){4}}
\put(56,55){x'L}

\put(-7,20){\vector(1,0){50}}
\put(-7,18){\vector(1,0){50}}
\put(-7,13){\vector(1,0){15}}
\put(-7,5){\vector(1,0){15}}
\put(0,16){\line(6,-1){48}}
\put(0,16){\line(-1,-6){3}}
\put(48,8){\line(-1,-6){3}}
\multiput(0,8)(4.8,0){10}{\line(1,0){3}}
\put(54,12){\vector(0,1){4}}
\put(54,12){\vector(0,-1){4}}
\put(56,11){x'L}
\put(-17,94){\Large a}
\put(-17,54){\Large b}
\put(-17,14){\Large c}

\put(-29,41){\vector(1,0){8}}
\put(-29,41){\vector(1,1){8}}
\put(-29,41){\vector(0,-1){8}}
\put(-27,32){$x$}
\put(-20,41){$s$}
\put(-20,46){$y$}

\end{picture}
\end{center}
\caption {
Horizontally tilted crystal:
(a) aligned,
(b) tilt $x'>$0,
(c) $x'<$0.
      }	\label{1}
\end{figure}
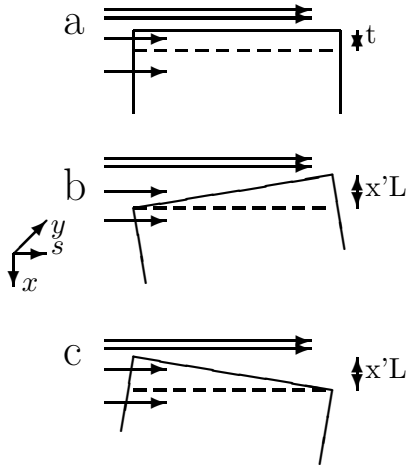

Consider a crystal disaligned from the
beam horizontally at $x'$.
Depending on the sign of $x'$, either the upstream end
approaches the beam (we define $x'<$0),
or the downstream one ($x'>$0); fig. 1.
Because of $x'$,
a {\em septum width} as thick as $t=|x'|L$
occurs at the crystal edge; $L$ is the crystal length (4 cm in E853).
Protons incident in the range 0$<$$b$$<t$ do not
traverse the full length of crystal.
The result depends dramatically on $x'$ sign.

In case of $x'>0$,
protons traverse the downstream edge.
It is disaligned by $\sim$0.64 mrad
(the bending angle) w.r.t. the beam.
Therefore, protons
traverse it like an amorphous substance.
This case imitates a crystal with an amorphous near-surface layer
as wide as $t\approx x'L$.
Measuring $F(x')$ for $x'$$>$0,
one measures $F(t)$.
Theory [9] predicts
very weak $F(t)$ dependence at high energies. The confirmation would be
encouraging for the multi-TeV crystal extraction.
Notice that the step of $t$ scan  could be very
fine: with $\delta x'$=2.5 $\mu$rad and $L$=4 cm one has
$\delta t$= 0.1 $\mu$m.

In case of $x'<0$,
protons traverse the upstream edge
which is aligned w.r.t. the beam.
Therefore, many particles are trapped in channeling.
However, those incident in the range 0$<$$b$$<$$x'L$ traverse a
reduced ($<$4 cm) length, thus getting a reduced ($<$0.64 mrad)
deflection and therefore are lost.
The inequality of two cases, $x'$$>$0 and $x'$$<$0,
causes a strong asymmetry of $F(x')$ dependence.
The difference $\Delta F$=$F(x')$--$F(-x')$
is proportional to the number of protons incident
with 0$<$$b$$<$$x'L$.
Varying $x'$ and observing $\Delta F$,
one investigates the distribution
over $b$ at crystal,
with accuracy of $\delta b$=0.1 $\mu$m.

This is complicated by another interesting phenomenon.
The protons incident on an aligned imperfect crystal
with $b_{max}<t$, have to traverse the full length of the crystal,
and to experience a substantial nuclear scattering.
Suppose, this crystal is disaligned
so that $b_{max}/x'L\approx$0.1. Then at first incidence the protons
traverse only the crystal edge, with the "length" $\le$0.1 that of crystal.
The respective scattering and losses
over 0.1$L$ are much smaller.
In this case
the protons retain better chances for extraction
with later passes than in the former case (perfect alignment).
The secondary $b$ of the scattered protons
are still sufficiently large ($\approx$30 $\mu$m $\gg$$x'L$ here),
so the "gap" $x'L$ is not dangerous.

We come to conclusion that a peak efficiency with {\em imperfect}
crystal is achieved at some tilt $x'\neq$0.
In the real experiment one scans $x'$ while
searching the peak, and comes to this case {\em automatically}.
We used the case $b_{max}/x'L=$0.1 as an illustration;
the optimal $x'$ will be found automatically in the scan.
Further on, we refer to this case as to the "pre-scatter" case,
when protons first gently pre-scatter in the crystal edge
to come later with low divergence but high $b$.
Understandably, with imperfect crystal the prescatter case may appear
also for a small negative tilt, $x'<$0.
Then, $F(x')$  may have {\em two} peaks, with a {\em dip}
at $x'$=0.
The width of the dip at $x'\approx$0
is also an indicator for $b_{max}$.

\section{Simulation procedure}

The crystal was located 61 m upstream of C0 point of Tevatron lattice,
with the edge at the horizontal distance of $X$=1.75 mm from the beam axis.
At the crystal location, the machine parameters were
$\beta_x$=105.7 m, $\alpha_x$=0.109 (horizontally), and
$\beta_y$=21.5 m, $\alpha_y$=0.148 (vertically); tunes
$Q_x$=20.5853 and $Q_y$=20.5744.
The beam invariant rms emittance was 2.5 mm$\cdot$mrad,
which corresponds to vertical rms divergence 11.5 $\mu$rad
and width 0.24 mm at the crystal location.

Crystal was a Si(110) slab 40 by 3 by 3 mm$^3$, 0.64 mrad bent,
with a perfect lattice,
and curved with a constant curvature
to deflect protons in vertical direction.
As an option, we model an amorphous layer at its edge
and/or irregularities of the surface.
The horizontal parameters $x,x'$ of incident particles
are defined by the mechanism of diffusion.
The two processes, diffusion and crystal extraction,
are unfold in E853.
Beam parameters in the channeling plane (vertical)
are not disturbed by the diffusion.
The exact value of $b_{max}$ matters only w.r.t. $t$.
Since $t$ is unknown for the real crystal,
we can postulate $b_{max}$=1 $\mu$m, and then model crystals
with different $t$.

\section{Results}

Fig. 2 shows the $F(y')$ angular scan for $x'$=0.
The peak $F$ of an ideal crystal is $\simeq$44 \%.
The same fig. shows scans for the crystals
with $t$=1 $\mu$m (i.e. $t$=$b_{max}$) and $t$=50 $\mu$m,
where at $y'$=$x'$=0
$F$$\approx$ 36 \% and 32 \% respectively.
However, for the imperfect crystals the real peak was found at $x'\neq$0
(Fig. 3).
With optimized $x'$, one has peak $F\approx$
42 \% and 35 \% for $t$=1 $\mu$m and $t$=50 $\mu$m respectively.
Notably, the efficiencies and scans are quite weakly
dependent on the crystal perfection.
The width of $y'$ scan was 50--55 $\mu$rad
{\em fwhm} in these cases.
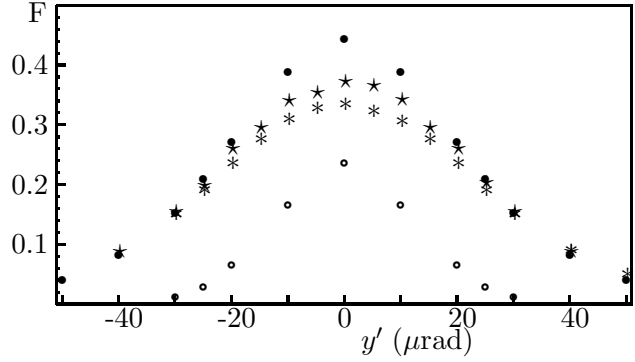
\begin{figure}[bth]
\begin{center}
\setlength{\unitlength}{.75mm}\thicklines
\begin{picture}(100,51)(-50,0)

\put(-51,0) {\line(1,0){102}}
\put(-51,0) {\line(0,1){53}}
\put(-51,53) {\line(1,0){102}}
\put( 51,0) {\line(0,1){53}}
\multiput(-50,0)(10,0){11}{\line(0,1){1.5}}
\multiput(-51,10.6)(0,10.6){4}{\line(1,0){1.5}}
\multiput(-51,0)(0,2.12){25}{\line(1,0){.5}}
\put(-40,-4){\makebox(2,1)[b]{-40}}
\put(-20,-4){\makebox(2,1)[b]{-20}}
\put(0,-4){\makebox(.2,.1)[b]{0}}
\put(20,-4){\makebox(2,1)[b]{20}}
\put(40,-4){\makebox(2,1)[b]{40}}
\put(-59,10.6){\makebox(.2,.1)[l]{0.1}}
\put(-59,21.2){\makebox(.2,.1)[l]{0.2}}
\put(-59,31.8){\makebox(.2,.1)[l]{0.3}}
\put(-59,42.4){\makebox(.2,.1)[l]{0.4}}

\put(-56,50){F}
\put(3,-8){$y'\; (\mu$rad)}

\put(10,41.2){\circle*{1.5}}
\put(20,28.8){\circle*{1.5}}
\put(25,22.2){\circle*{1.5}}
\put(30,16.2){\circle*{1.5}}
\put(40,8.7){\circle*{1.5}}
\put(50,4.3){\circle*{1.5}}
\put(0,47){\circle*{1.5}}
\put(-10,41.2){\circle*{1.5}}
\put(-20,28.8){\circle*{1.5}}
\put(-25,22.2){\circle*{1.5}}
\put(-30,16.2){\circle*{1.5}}
\put(-40,8.7){\circle*{1.5}}
\put(-50,4.3){\circle*{1.5}}

\put(0,25){\circle{1}}
\put(10,17.6){\circle{1}}
\put(20,7){\circle{1}}
\put(25,3.1){\circle{1}}
\put(30,1.2){\circle{1}}
\put(-10,17.6){\circle{1}}
\put(-20,7){\circle{1}}
\put(-25,3.1){\circle{1}}
\put(-30,1.2){\circle{1}}

\put(-1,34.3){$\ast$}
\put(-6,33.6){$\ast$}
\put(-11,31.6){$\ast$}
\put(-16,28.1){$\ast$}
\put(-21,23.7){$\ast$}
\put(-26,19){$\ast$}
\put(-31,14.8){$\ast$}
\put(4,33.1){$\ast$}
\put(9,31.3){$\ast$}
\put(14,28){$\ast$}
\put(19,23.7){$\ast$}
\put(24,18.9){$\ast$}
\put(29,14.8){$\ast$}
\put(39,8.4){$\ast$}
\put(49,4){$\ast$}

\put(-1,38.4){$\star$}
\put(-6,36.4){$\star$}
\put(4,37.6){$\star$}
\put(-11,34.9){$\star$}
\put(-16,30.2){$\star$}
\put(-21,26.4){$\star$}
\put(-26,19.8){$\star$}
\put(-31,15.3){$\star$}
\put(-41,8.2){$\star$}
\put(9,35.1){$\star$}
\put(14,30.2){$\star$}
\put(19,26.4){$\star$}
\put(24,20.5){$\star$}
\put(29,15.3){$\star$}
\put(39,8.2){$\star$}

\end{picture}
\end{center}
\caption { $F(y')$ scan for $x'$=0.
Ideal crystal: (o) first-pass
and ($\bullet$) overall efficiencies.
Imperfect crystal:
overall efficiency with $t$=1 $\mu$m ($\star$)
and $t$=50 $\mu$m ($\ast$).
      }	\label{r1}
\end{figure}

The angular scan $F(x')$ is in Fig. 3.
The depth of the dip at $x'\approx$0 (i.e. at perfect alignment)
is $\simeq$14 \% and $\simeq$7 \% w.r.t. the peak
for $t$=1 $\mu$m and $t$=50 $\mu$m respectively.
The width $\Delta x'$ of the peculiarity
(either peak or dip) near $x'\approx$0
is roughly $b_{max}/L$ which is 25 $\mu$rad in our simulation.
\begin{figure}[bth]
\begin{center}
\setlength{\unitlength}{1.0mm}\thicklines
\begin{picture}(50,53)(-17,0)

\put(-18,0) {\line(1,0){50}}
\put(-18,0) {\line(0,1){53}}
\put(-18,53) {\line(1,0){50}}
\put( 32,0) {\line(0,1){53}}
\multiput(-15,0)(5,0){10}{\line(0,1){1.5}}
\multiput(-18,10.6)(0,10.6){4}{\line(1,0){1.5}}
\multiput(-18,0)(0,2.12){25}{\line(1,0){.5}}
\put(-10,-4){\makebox(2,1)[b]{-100}}
\put(0,-4){\makebox(.2,.1)[b]{0}}
\put(10,-4){\makebox(2,1)[b]{100}}
\put(20,-4){\makebox(2,1)[b]{200}}
\put(-24,10.6){\makebox(.2,.1)[l]{0.1}}
\put(-24,21.2){\makebox(.2,.1)[l]{0.2}}
\put(-24,31.8){\makebox(.2,.1)[l]{0.3}}
\put(-24,42.4){\makebox(.2,.1)[l]{0.4}}

\put(-23,50){F}
\put(1,-7){$x'$ ($\mu$rad)}

\put(1.5,43.9){$\bullet$}
\put(2.5,42){$\bullet$}
\put(5,43.6){$\bullet$}
\put(10,45.1){$\bullet$}
\put(17.5,44.6){$\bullet$}
\put(25,43.7){$\bullet$}

\put(0,47){$\bullet$}
\put(-.5,43.9){$\bullet$}
\put(-1,40.2){$\bullet$}
\put(-1.5,36){$\bullet$}
\put(-2,32.7){$\bullet$}
\put(-2.5,30){$\bullet$}
\put(-5,29.1){$\bullet$}
\put(-10,27.8){$\bullet$}

\put(0,34.3){$\ast$}
\put(.5,34.8){$\ast$}
\put(1,35.5){$\ast$}
\put(2,36){$\ast$}
\put(5,36.9){$\ast$}
\put(7.5,36.4){$\ast$}
\put(10,36.8){$\ast$}
\put(15,36){$\ast$}
\put(20,35.4){$\ast$}

\put(-.5,34.3){$\ast$}
\put(-1,35){$\ast$}
\put(-2,35.8){$\ast$}
\put(-5,36.2){$\ast$}
\put(-7.5,35.5){$\ast$}
\put(-10,36.9){$\ast$}
\put(-15,35.4){$\ast$}

\put(0,38.4){$\star$}
\put(.5,38.2){$\star$}
\put(1,39.5){$\star$}
\put(2,39.7){$\star$}
\put(5,43){$\star$}
\put(10,44.1){$\star$}
\put(15,43.2){$\star$}
\put(20,44.7){$\star$}
\put(25,43.1){$\star$}
\put(30,42.6){$\star$}

\put(-.5,38.7){$\star$}
\put(-2,41){$\star$}
\put(-5,42.8){$\star$}
\put(-7,41.5){$\star$}
\put(-10,40.6){$\star$}
\put(-15,38.9){$\star$}

\end{picture}
\end{center}
\caption { $F(x')$ scan near the peak;
see fig.2 caption.
      }	\label{r3}
\end{figure}
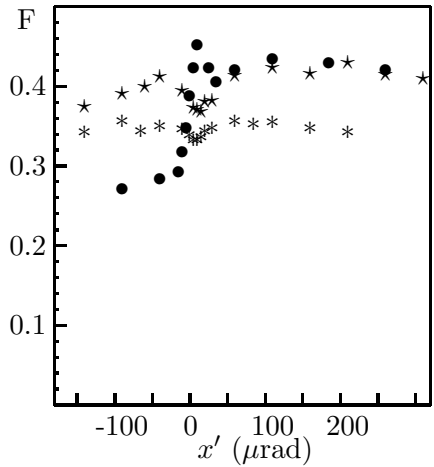
 $F$ is 1/2 of the maximum at $x'\simeq$14 mrad and
 $-$0.3 mrad  for an ideal crystal ({\em fwhm}
 of the horizontal scan is $\simeq$14 mrad),
 at $x'\simeq$15 mrad and --1.2 mrad
 for the crystal with $t$=1 $\mu$m ({\em fwhm}$\simeq$16 mrad),
 and at $x'\simeq$18 mrad and --5 mrad
 for the crystal with $t$=50 $\mu$m ({\em fwhm}$\simeq$23 mrad).

The asymmetry of the scan, $F(x')\neq F(-x')$, is due to
the loss of the protons trapped in channeling
near the crystal edge.
With an ideal crystal, the asymmetry exists for any $x'$.
With a septum width $t$, the asymmetry can be seen for an
angling $\pm x'$ larger than $t/L$ only.
In our simulation with $t$=50 $\mu$m,
the scan is symmetric indeed within $\pm$1.3 mrad but asymmetric
outside this range of $x'$; note that 50 $\mu$m/40 mm =1.25 mrad.
We expect therefore this $x'$-threshold for an asymmetry  to be
a good measure of the septum width $t$.
If one plots the magnitude of asymmetry,
$F(x')$--$F(-x')$, as a function of $x'L$, he obtains a rough estimate
of the beam distribution over $b$ at crystal.
The minimal step $\delta b=\delta x'L$=0.1 $\mu$m
is much finer than the precision of coordinate detectors
$\approx$0.1 mm !

Notice an abrupt decrease in $F$ of the ideal crystal
over the range of $x'L$ from 0 to $-b_{max}$:
from 44 \% at $x'$=0 to 28 \% at $x'$= $-b_{max}/L$.
This drop is an excellent opportunity to measure the primary $b_{max}$
with a precision of $\delta b$=0.1 $\mu$m.
With an ideal crystal one can measure
a distribution over the {\em primary}
$b$ (in the range of $\sim$1 $\mu$m).
With imperfect crystal
in the same way one measures the
distribution over {\em secondary}
$b$ (in a broad range from $\sim$$t$ to $\sim$1 mm).
Finally, the dependence $F(x')$ for $x'$$>$0 gives actually
the dependence of $F$ on the septum width
$t\simeq x'L$.

The distribution of the extracted particles
over $x$ at the crystal face
is  essential
for understanding both the crystal interplay with
other accelerator elements and the requirements for
the crystal face perfection.
We have found that one half of the extracted protons have penetrated
into the crystal depth by $>$0.3 mm; another half had $b$$<$0.3 mm.
The $y'$ divergence of the extracted beam
 was defined by the channeling properties of Si(110) crystal;
 its full width 2$\theta_c$ was $\approx$12.8 $\mu$rad
 ($\theta_c$$\simeq$6.4 $\mu$rad),
 and {\em fwhm}$\approx$9 $\mu$rad.
The $x'$ divergence was $\simeq$5 $\mu$rad {\em fwhm}
with the ideal crystal and $\sim$12 $\mu$rad {\em fwhm} with $t$=1 $\mu$m.
It was increased due to scattering in inefficient passes.
After the doublet of quadrupoles and
the Lambert\-son-type magnet,
two detectors (hodoscopes with 0.1 mm bins)
were placed at 80.5 m (D1)
and 120.5 m (D2) downstream of the crystal
to measure the bent-beam profiles.
The horizontal profiles had
width $\simeq$ 0.3 and 0.4 mm {\em fwhm}
for the ideal and $t$=1 $\mu$m crystals
at D1, and $\simeq$ 0.5 and 0.7--0.9 mm {\em fwhm}
at D2.

\section{Optimization}

The extraction efficiency $F$ is defined by the processes
of channeling, scattering, and nuclear interaction in crystal.
All the processes depend essentially on the crystal length $L$.
Fig. 4 shows $F(L)$.
$F$ is maximal, near 70 \%,
in $L$ range 0.4--1.0 cm, irrespective
of the crystal perfection.
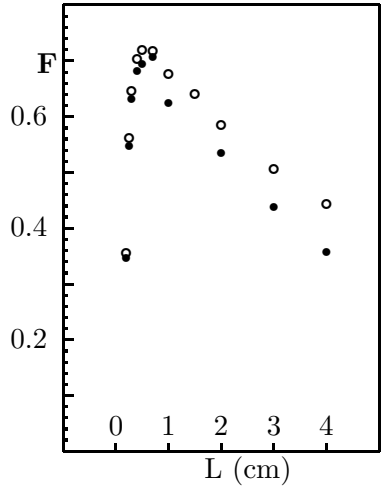
\begin{figure}
\begin{center}
\setlength{\unitlength}{.7mm}
\begin{picture}(80,85)(-20,0)
\thicklines

\put(2,37.7){\circle{1.5}}
\put(2.5,59.6){\circle{1.5}}
\put(3,68.4){\circle{1.5}}
\put(4,74.5){\circle{1.5}}
\put(5,76.4){\circle{1.5}}
\put(7,76.0){\circle{1.5}}
\put(10,71.7){\circle{1.5}}
\put(15,67.9){\circle{1.5}}
\put(20,62.){\circle{1.5}}
\put(30,53.6){\circle{1.5}}
\put(40,47.0){\circle{1.5}}

\put(2,36.7){\circle*{1.5}}
\put(2.5,58.1){\circle*{1.5}}
\put(3,67.){\circle*{1.5}}
\put(4,72.4){\circle*{1.5}}
\put(5,73.6){\circle*{1.5}}
\put(7,75.0){\circle*{1.5}}
\put(10,66.3){\circle*{1.5}}
\put(20,56.7){\circle*{1.5}}
\put(30,46.5){\circle*{1.5}}
\put(40,38){\circle*{1.5}}

\put(-10,0) {\line(1,0){60}}
\put(-10,0) {\line(0,1){85}}
\put(-10,85) {\line(1,0){60}}
\put(50,0){\line(0,1){85}}
\multiput(0,0)(10,0){6}{\line(0,1){1.5}}
\put(-.5,3.){\makebox(1,.5)[b]{0}}
\put(9.5,3.){\makebox(1,.5)[b]{1}}
\put(19.5,3.){\makebox(1,.5)[b]{2}}
\put(29.5,3.){\makebox(1,.5)[b]{3}}
\put(39.5,3.){\makebox(1,.5)[b]{4}}
\multiput(-10,0)(0,10.6){8}{\line(1,0){2}}
\multiput(-10,0)(0,2.12){40}{\line(1,0){.75}}
\put(-20,21.2){\makebox(1,.5)[l]{0.2}}
\put(-20,42.4){\makebox(1,.5)[l]{0.4}}
\put(-20,63.6){\makebox(1,.5)[l]{0.6}}

\put(-15,72){\bf F}
\put(15,-5){ L (cm)}

\end{picture}
\caption {$F(L)$ for ideal (o) and
 $t$=1 $\mu$m ($\bullet$) crystals.
      }	\label{fl}
\end{center}
\end{figure}

\section{Conclusions}

The key information of the
multi-pass crystal extraction
can be obtained from the analysis of the horizontal
angular scan of efficiency.
In the considered way
one can study the impact parameters of halo particles
and the structure of the crystal edge with
an accuracy as fine as 0.1 $\mu$m.

The extraction efficiency is expected as high as $\simeq$40 \%
irrespective of the crystal septum width,
and can be increased up to $\sim$70 \%
with the use of a shorter ($\le$1 cm) crystal.
The difference in efficiency between the ideal and imperfect crystals
is very low, because of predominance of the multi-passes in extraction
at high energies,
and partly due to the found effect of a gentle "prescattering"
in the edge of a crystal tilted horizontally.
This provides an elementary solution to the problem
of a finite septum width and infinitesimal impact parameters.

One general trend in the results of simulations,
from SPS [7] to Tevatron to LHC [6],
is worthwhile to mention:
the difference in efficiency of the ideal crystal and crystal
with imperfect surface vanishes with energy $E$,
because the scattering angle reduces faster ($\sim$1/$E$)
than $\theta_c$ does ($\sim$1/$\sqrt{E}$).


\end{document}